\documentstyle[preprint,aps]{revtex}
\tighten
\begin{document}
\draft
\preprint{\vbox{Submitted to Physical Review C \hfill FSU-SCRI-95-81 \\
                                          \null\hfill IU/NTC 95-10  \\
                                          \null\hfill UK/95-08  \\
                                          \null\hfill nucl-th/9508035}}
\title{Isospin-Violating Meson-Nucleon Vertices as an \\
Alternate Mechanism of Charge-Symmetry Breaking}
\author{S. Gardner}
\address{Department of Physics and Astronomy, \\
         University of Kentucky, Lexington, KY 40506}
\author{C.~J. Horowitz}
\address{Nuclear Theory Center and Department of Physics, \\
         Indiana University, Bloomington, IN 47505}
\author{J. Piekarewicz}
\address{Supercomputer Computations Research Institute, \\
         Florida State University, Tallahassee, FL 32306}
\date{\today}
\maketitle
 \begin{abstract}
We compute isospin-violating meson-nucleon coupling
constants and their consequent charge-symmetry-breaking
nucleon-nucleon potentials. The couplings result from
evaluating matrix elements of quark currents between
nucleon states in a nonrelativistic constituent quark
model; the isospin violations arise from the difference
in the up and down constituent quark masses. We find, in particular,
that isospin violation in the omega-meson--nucleon vertex dominates
the class IV CSB potential obtained from these considerations.
We evaluate the resulting spin-singlet--triplet mixing angles,
the quantities germane to the difference of neutron and proton
analyzing powers measured in elastic $\vec{n}-\vec{p}$ scattering,
and find them commensurate to those computed originally
using the on-shell value of the $\rho$-$\omega$ mixing amplitude.
The use of the on-shell $\rho$-$\omega$ mixing amplitude at $q^2=0$
has been called into question; rather, the amplitude is zero in a
wide class of models. Our model possesses no contribution from
$\rho$-$\omega$ mixing at $q^2=0$, and we find that omega-meson
exchange suffices to explain the measured $n-p$ analyzing power
difference~at~183 MeV.
\end{abstract}

\narrowtext

\section{Introduction}
\label{secintro}

  The suggestion of Goldman, Henderson, and Thomas~\cite{ght92} that the
contribution of $\rho$-$\omega$ mixing to charge-symmetry-breaking
(CSB) observables is suppressed in the low momentum transfer regime
has opened the search for new sources of isospin violation. Since
then, many calculations, using a variety of models, have confirmed the
suppression of the $\rho$-$\omega$ mixing amplitude at small spacelike
momenta~\cite{piewil93,hats93,krein93,mitch94,oconn94,maltman95}.
Indeed, it has been shown that the $\rho$-$\omega$ mixing amplitude is
zero at $q^2=0$ in all models with vector mesons coupled to conserved
currents~\cite{oconn94}. Yet, in
Refs.~\cite{piewil93,hats93,krein93,mitch94,oconn94,maltman95} no
alternate mechanisms to $\rho$-$\omega$ mixing are proposed.
The phenomenological impact of this gap must be emphasized:
the CSB potential from $\rho$-$\omega$ mixing --- with the mixing amplitude
fixed at the omega-meson point --- accounts for some 40\% of
the difference between the neutron and proton analyzing powers ($\Delta A$)
measured in elastic $\vec{n}-\vec{p}$ scattering at 183 MeV~\cite{knut90}.
Without this contribution the previous agreement between theory and
experiment would be upset~\cite{miller86,willia87,miller90,iqnisk94}. Although
the suppression of the mixing amplitude continues to be
controversial~\cite{miller94,oconn95}, sources of additional isospin
violation are interesting in their own right and deserve examination.
Indeed, the aim of the present paper is to show that a recently
proposed CSB mechanism --- based on isospin-violating meson-nucleon
coupling constants~\cite{ghp95} --- is sufficient to restore
the agreement with experiment. Specifically, we examine the effect
of these new sources of CSB on the spin-singlet--triplet mixing
angles; these are the fundamental dynamical quantities driving
$\Delta A$~\cite{willia87}.

  Most theoretical efforts devoted to understanding CSB
observables use a nucleon-nucleon ($NN$) interaction
constrained by two-nucleon
data~\cite{miller86,willia87,holz87}. In such a picture, isospin
violations arise from electromagnetic effects and hadronic mass
differences. Sources of CSB can be classified in terms of three
distinct contributions: (i) isovector-isoscalar mixing in the
meson propagator, (ii) isospin-breaking in the nucleon wave
function, and (iii) isospin breaking in the meson-nucleon and
photon-nucleon vertices. Rho-omega mixing, the
proton-neutron mass difference, and the difference between
the electric charge of the proton and the neutron are typical
examples of (i), (ii), and (iii), respectively.
 The existence of isovector-isoscalar mixing, such as $\pi$-$\eta$ and
$\rho$-$\omega$ mixing, is well established. For example,
$\rho$-$\omega$ mixing has been observed experimentally in
$e^{+}e^{-}\rightarrow\pi^{+}\pi^{-}$ measurements at the
$\omega-$meson production point~\cite{barkov85}. However,
the suggested suppression of the mixing amplitudes at small
spacelike momenta lessens their impact on
CSB observables. This is, in part, why we consider other
sources of isospin violation in this paper.

  Isospin breaking in the nucleon wave function in a hadronic model
is driven by the neutron-proton mass difference. Indeed, it is through
this mechanism that charged-pion exchange dominates~\cite{cheung80,gersten81}
the class IV potential~\cite{henmil79} at moderate momentum transfers.
Isospin breaking in the nucleon wave function can also arise in a quark
model picture from the mixing of the nucleon to
$|J^{\pi}=1/2^{+};T=3/2\rangle$ baryon states~\cite{dmitra95}. While
undoubtedly nonzero, one expects the $T=3/2$ components of the nucleon
to be small due to the large mass difference between the nucleon and
the $\Delta(1910)$ --- the first $P_{31}$ baryon. In contrast, the
$\rho$-$\omega$ mass difference is a mere 12 MeV. Thus, we turn to
the meson-nucleon coupling constants as the possible sources of
isospin violation demanded by data.

  While there have been calculations of isospin-violating meson-nucleon
coupling constants~\cite{miller90,mitra67,thomas81,henzha87,iqnisk88}, their
impact on class IV CSB observables has only recently been
considered~\cite{ghp95}. Here, as earlier~\cite{ghp95}, we adopt a
nonrelativistic quark model to calculate isospin breaking in the
meson-nucleon coupling constants. In the model the coupling constants
emerge from evaluating matrix elements of quark currents of the
appropriate Lorentz and flavor structure between nucleon states. The
isospin violations arise from the up-down quark mass difference.
Radiative corrections to the vertices have also been evaluated
and are found to be small~\cite{morris68,yakin79}. Here we study
the phenomenological impact of Ref.~\cite{ghp95} in greater detail.
In order to do this, we estimate the $q^2$ dependence of the
isospin-breaking found in the vertices at $q^2=0$.

  We have organized the paper as follows. In Sec.~\ref{secivcc}
the model is introduced, and isospin-violating meson-nucleon
coupling constants are computed. We show that in the $q^2=0$ limit
the couplings depend merely on the spin-flavor structure of the
nucleon wave function; they are insensitive to the spatial components
of the nucleon wave function. In Sec.~\ref{seccsbp} we use these
findings to compute the resulting CSB potentials. In particular, we
obtain a large contribution from omega-meson exchange to the class IV
potential. We quantify the impact of isospin-violation in the
$NN\omega$ vertex by computing the resulting spin-singlet--triplet
mixing angles --- these are the basic building blocks of $\Delta A$.
These results are presented in Sec.~\ref{secresults}. Finally, we
discuss the impact of our work in Sec.~\ref{secconcl}.

\section{Isospin-violating meson-nucleon coupling constants}
\label{secivcc}

We are interested in computing the coupling of an on-shell nucleon
to the neutral mesons $\omega$, $\rho^{0}$, $\pi^{0}$, and $\sigma$.
The off-shell vertices could engender additional isospin breaking,
but our primary focus is on the $NN$ system, so that we will not
consider these effects further. The exchanged mesons couple to nucleon
currents of the appropriate Lorentz character, and the meson-nucleon
coupling constants emerge from evaluating the matrix elements of these
currents in the quark model. The difference in the up and down constituent
quark masses thus gives rise to isospin-violating meson-nucleon coupling
constants. At $q^{2}=0$ these couplings are determined from the spin and
flavor structure of the nucleon wave function alone. In contrast, the
couplings at $q^2=0$ of the nucleon to the charged mesons are sensitive
to the quark momentum distribution as well, and are, therefore, more
model dependent~\cite{miller90}. We shall consider the
neutral-vector-meson--nucleon vertices first, as they are relevant to
the $\Delta A$ measurement. The most general form for these on-shell
$NN$-meson vertex functions, consistent with Lorentz covariance
and parity invariance, are
\begin{mathletters}
\label{vert}
\begin{eqnarray}
   -ig_{\lower 2pt \hbox{$\scriptstyle NN\omega$}}
   \Lambda^{\mu}_{\lower 2pt \hbox{$\scriptstyle NN\omega$}} &=&
   -ig_{\lower 2pt \hbox{$\scriptstyle NN\omega$}}
    \left(
      g^{\omega}_N \gamma^{\mu} +
     if^{\omega}_N \sigma^{\mu\nu} {(p'-p)_{\nu} \over 2M_{N}}
     \right) \;, \\
   -ig_{\lower 2pt \hbox{$\scriptstyle NN\rho$}}
   \Lambda^{\mu}_{\lower 2pt \hbox{$\scriptstyle NN\rho$}} &=&
   -ig_{\lower 2pt \hbox{$\scriptstyle NN\rho$}}
    \left(
      g^{\rho}_N \gamma^{\mu} +
     if^{\rho}_N \sigma^{\mu\nu} {(p'-p)_{\nu} \over 2M_{N}}
     \right) \;,
 \end{eqnarray}
\end{mathletters}
where $g_{\lower 2pt \hbox{$\scriptstyle NN\alpha$}}$
($\alpha=\omega,\rho$) are the isospin-averaged, phenomenological,
meson-nucleon coupling constants, determined from
fits to the $NN$ phase shifts and to the properties of the
deuteron~\cite{machl87,machl89}, and $M_{N}$ is the nucleon mass.
We compute the couplings, that is, $g^{\alpha}_{N}(q^2)$ and
$f^{\alpha}_{N}(q^2)$, by assuming that the $NN\alpha$ vertex functions
can be related to the matrix elements of quark currents of the
appropriate Lorentz and flavor structure between nucleon states,
computed in the nonrelativistic quark model. Thus,
\begin{mathletters}
 \begin{eqnarray}
  \langle N(p',s') | J^{\mu;+} | N(p,s) \rangle &=&
  \bar{U}(p',s')
   \Lambda^{\mu}_{\lower 2pt \hbox{$\scriptstyle NN\omega$}}
  U(p,s) \;, \\
  \langle N(p',s') | J^{\mu;-} | N(p,s) \rangle &=&
  \bar{U}(p',s')
   \Lambda^{\mu}_{\lower 2pt \hbox{$\scriptstyle NN\rho$}}
  U(p,s) \;.
 \end{eqnarray}
 \label{vertex}
\end{mathletters}
Here $U(p,s)$ denotes a on-shell nucleon spinor of mass $M_{N}$,
momentum $p$, and spin $s$. We shall focus on the couplings at
$q^2=0$, where $q\equiv p'-p$, as the nonrelativistic quark model
is best suited to an estimate in the static limit. The quark currents
$J^{\mu;\pm}$ are
\begin{mathletters}
 \begin{eqnarray}
  J^{\mu;+}   &=&   {1\over 3}\, \bar{u} \gamma^{\mu} u
               +    {1\over 3}\, \bar{d} \gamma^{\mu} d \;, \\
  J^{\mu;-}   &=&                \bar{u} \gamma^{\mu} u
               -                 \bar{d} \gamma^{\mu} d \;.
 \end{eqnarray}
 \label{veccurrent}
\end{mathletters}
It is the quark vector current which is appropriate to the
vector-meson--nucleon vertex; the second superscript ($\pm$)
denotes its symmetry under the $u \leftrightarrow d$ flavor
transformation. Note that the constituent quarks are assumed
to be elementary: no quark form factors have been introduced.
The isoscalar vector quark charge is $1/3$, whereas
the isovector vector quark charge is $+1$ for the up quark and
$-1$ for the down quark. The charge assignments are made such
that $g_N^{\omega}=1$, $g_p^{\rho}=1$, and $g_n^{\rho}=-1$,
at $q^2=0$.

Our model stems from the notion of vector dominance~\cite{nambu57}.
Vector dominance presumes that a photon's interaction with a nucleon
is mediated by the rho --- or omega --- meson. Here we argue that
the coupling of the vector mesons themselves to the nucleon can be
determined via matrix elements of the appropriate isospin
components of the quark {\it vector} current.
Our model does not predict the isospin-conserving coupling
constants $g_{\lower 2pt \hbox{$\scriptstyle NN\alpha$}}$; these must
be extracted from phenomenological fits to two-nucleon data. However,
the isospin-violating pieces, as well as the
tensor-to-vector ratio, can be calculated within the model.
Note that the vector dominance nature of our model implies that the
quarks couple to {\it conserved currents}. We estimate the resulting
coupling constants using the nonrelativistic quark model (NRQM); this
is an additional assumption.

The couplings $g^{\alpha}_{N}$ and $f^{\alpha}_{N}$ are
functions of the meson four-momentum $q^{2}$,
though we shall focus on the couplings at
$q^2=0$. In this limit the couplings are
insensitive to the spatial component of the nucleon wave function; they
follow directly from its spin and flavor content alone.
In the $SU(6)$ limit~\cite{perkins87},
\begin{eqnarray}
  |p \!\uparrow \rangle = {1 \over \sqrt{18}}
  \Big(
   && 2|u\!\uparrow u\!\uparrow d\!\downarrow\rangle -
       |u\!\uparrow u\!\downarrow d\!\uparrow\rangle -
       |u\!\downarrow u\!\uparrow d\!\uparrow\rangle +    \nonumber \\
   && 2|u\!\uparrow d\!\downarrow u\!\uparrow\rangle -
       |u\!\downarrow d\!\uparrow u\!\uparrow\rangle -
       |u\!\uparrow d\!\uparrow u\!\downarrow\rangle +    \\
   && 2|d\!\downarrow u\!\uparrow u\!\uparrow\rangle -
       |d\!\uparrow u\!\uparrow u\!\downarrow\rangle -
       |d\!\uparrow u\!\downarrow u\!\uparrow\rangle     \nonumber
   \Big) \;.
\end{eqnarray}
The neutron spin-up wave function, $|n\!\uparrow \rangle$, is obtained
by exchanging the up and down quarks in the expression for
$|p \!\uparrow \rangle$. The isospin violations arise from the
difference in the up and down constituent quark masses. The couplings
constants are obtained from computing the matrix elements found in the
nonrelativistic reduction of Eq.~(\ref{vertex}) in the quark model;
{\it i.e.},
\begin{mathletters}
\label{themodel}
 \begin{eqnarray}
    g^{\omega}_N  = \sum_{i=1}^3
    g^{+}_i
    \langle N \!\uparrow | 1 | N\! \uparrow\rangle  \quad &;& \quad
   {(g^{\omega}_N + f^{\omega}_N) \over {2M_N}} = \sum_{i=1}^3
    \mu^{+}_i
    \langle N \!\uparrow | \sigma^z_i | N\! \uparrow\rangle   \;, \\
    g^{\rho}_N  = \sum_{i=1}^3
    g^{-}_i
    \langle N \!\uparrow | 1 | N\! \uparrow\rangle  \quad &;& \quad
   {(g^{\rho}_N + f^{\rho}_N) \over {2M_N}} = \sum_{i=1}^3
   \mu^{-}_i
   \langle N \!\uparrow | \sigma^z_i | N\! \uparrow\rangle   \;.
 \end{eqnarray}
\end{mathletters}
Note that we have introduced the quark magnetic moment
$\mu^{\pm}_{i} \equiv g^{\pm}_{i}/2m_{i}$, with the
charges $g^{\pm}_{i}$ given in Table~\ref{tableone}.
In the following presentation we discuss only the coupling
of the nucleon to the $\omega-$meson, as an illustrative
example. Our results, collected in Table~\ref{tabletwo},
include the couplings to the other mesons as well.

The vector coupling of the $\omega-$meson to the nucleon is
determined by simply counting the quark charges:
\begin{mathletters}
 \begin{eqnarray}
  g^{\omega}_p &=&
  2g^{+}_u + g^{+}_d = 1 \;, \\
  g^{\omega}_n &=&
  2g^{+}_d + g^{+}_u = 1 \;.
 \end{eqnarray}
 \label{model1}
\end{mathletters}
The tensor coupling, in contrast, depends on the spin structure
of the nucleon wave function:
\begin{mathletters}
 \begin{eqnarray}
   \mu^{\omega}_p \equiv
   {g^{\omega}_p + f^{\omega}_p \over {2M_p}} &=&
   {4 \over 3} \mu^{+}_u - {1 \over 3} \mu^{+}_d =
   {1 \over 18} \left(
   {4 \over m_{u}} - {1 \over m_{d}} \right) \;, \\
   \mu^{\omega}_n \equiv
   {g^{\omega}_n + f^{\omega}_n \over {2M_n}} &=&
   {4 \over 3} \mu^{+}_d - {1 \over 3} \mu^{+}_u =
   {1 \over 18} \left(
   {4 \over m_{d}} - {1 \over m_{u}} \right) \;.
 \end{eqnarray}
 \label{model2}
\end{mathletters}
It is
useful to construct
isoscalar and isovector combinations at the {\it nucleon} level; i.e.,
\begin{eqnarray}
    g^{\omega}_N &=&
     g^{\omega}_p {1 \over 2}(1+\tau_z) +
     g^{\omega}_n {1 \over 2}(1-\tau_z) \equiv
     g^{\omega}_0 + g^{\omega}_1 \tau_z \;,  \\
   \mu^{\omega}_N &=&
    \mu^{\omega}_p {1 \over 2}(1+\tau_z) +
    \mu^{\omega}_n {1 \over 2}(1-\tau_z) \equiv
    \mu^{\omega}_0 + \mu^{\omega}_1 \tau_z \;,
\end{eqnarray}
where
\begin{eqnarray}
\label{omegag}
    g^{\omega}_0 +   g^{\omega}_1 \tau_z    &=&     1 \;, \\
  \mu^{\omega}_0 + \mu^{\omega}_1 \tau_z    &=&
\label{omegaf}
  {1 \over 6m} \left[ 1 + {5\over 6}{\Delta m \over m} \tau_z \right] \equiv
  \left[ {(g^{\omega}_0 + f^{\omega}_0) \over 2M} +
        {(g^{\omega}_1 + f^{\omega}_1) \over 2M} \tau_z \right]
\end{eqnarray}
with
\begin{equation}
    M \equiv {1 \over 2}(M_{n}+M_{p}) \;; \quad
    m \equiv {1 \over 2}(m_{d}+m_{u}) \;; \quad
    \Delta m \equiv (m_{d}-m_{u}) \;.
\end{equation}
The expression in Eq.~(\ref{omegaf}) is given to leading order in
$\Delta m /m$ only. Note that isospin breaking in the $f$ and $g$
couplings is realized in the $f$ {\it alone} and that the breaking
in the $\omega$ tensor coupling is isovector in character. The
$\omega$ --- and $\rho$ --- vector couplings are isospin-conserving.
The isospin-breaking in our model is connected to that of the
electromagnetic form factors through our assumption of vector
dominance; charge conservation protects the charge form factor
from isospin-breaking at zero momentum transfer~\cite{dmitra95}.
The tensor coupling is explicitly sensitive to the quark mass,
as seen in Eq.~(\ref{themodel}), and the isospin-breaking
corrections are generated by the up-down mass difference.
In the constituent quark model $\Delta m > 0$~\cite{licht89};
the up quark, which is lighter, has a larger magnetic moment than
the down quark. Henceforth we shall adopt the choice $M/3m \equiv 1$
in reporting the coupling constants. Our results are summarized
in Table~\ref{tabletwo}.

We now consider the isospin-conserving results. We find for the
tensor-to-vector ratio that
\begin{equation}
{f^{\omega}_{0} \over g^{\omega}_N} =0 \;; \quad
{f^{\rho}_{1} \over g^{\rho}_N} =4 \;.
\end{equation}
These results are qualitatively consistent with the $f^V_N/g^V_N$
ratios which emerge from phenomenological fits to the $NN$
interaction~\cite{machl87,machl89} --- recall that the Bonn B
potential parameters~\cite{machl89}, for example, are
$f^{\omega}_N/g^{\omega}_N=0$  and $f^{\rho}_N/g^{\rho}_N=6.1$.
This consistency is intimately connected to the
NRQM's ability to describe the nucleon magnetic
moments and to our assumption of vector dominance. In the NRQM,
with $M=3m$, the anomalous magnetic moment is purely isovector:
$\kappa_p=2$ and $\kappa_n=-2$. Note that $\kappa_p^{\rm exp}=1.79$
and $\kappa_n^{\rm exp}=-1.91$. These successes gives us confidence
in using our model to compute the isospin-violating corrections to
these coupling constants.

  For completeness, we shall now consider isospin breaking in the
$NN\sigma$ and $NN\pi^0$ vertices as well. We exclude the $NN\eta$
vertex from this discussion because the
$g_{\lower 2pt \hbox{$\scriptstyle NN\eta$}}$ coupling constant
is poorly constrained by $NN$ data~\cite{coon82}. The appropriate
vertex functions are
\begin{mathletters}
\begin{eqnarray}
\label{vertsig}
    ig_{\lower 2pt \hbox{$\scriptstyle NN\sigma$}}
   \Lambda^{\rm s}_{\lower 2pt \hbox{$\scriptstyle NN\sigma$}} &=&
    ig_{\lower 2pt \hbox{$\scriptstyle NN\sigma$}}
    \left(
      g^{\sigma}_N
    \right) {1} \;, \\
     g_{\lower 2pt \hbox{$\scriptstyle NN\pi$}}
    \Lambda^{5}_{\lower 2pt \hbox{$\scriptstyle NN\pi$}} &=&
     \phantom{i}
     g_{\lower 2pt \hbox{$\scriptstyle NN\pi$}}
    \left(
      g^{\pi}_N
    \right)\gamma^{5}  \;.
 \end{eqnarray}
\end{mathletters}
We have assumed pseudoscalar, rather than pseudovector, coupling
for the pion in order to be consistent with earlier calculations
of charge-symmetry
breaking~\cite{miller86,willia87,holz87,cheung80,gersten81}.
In our model, we connect the vertex functions
to matrix elements of quark currents, so that
\begin{mathletters}
 \begin{eqnarray}
  \langle N(p',s') | J^{\rm s} | N(p,s) \rangle &=&
  \bar{U}(p',s')
   \Lambda^{\rm s}_{\lower 2pt \hbox{$\scriptstyle NN\sigma$}}
  U(p,s) \;, \\
  \langle N(p',s') | J^{5} | N(p,s) \rangle &=&
  \bar{U}(p',s')
   \Lambda^{5}_{\lower 2pt \hbox{$\scriptstyle NN\pi$}}
  U(p,s) \;,
 \end{eqnarray}
 \label{vertexsig}
\end{mathletters}
where
\begin{mathletters}
 \begin{eqnarray}
  J^{\rm s}(q) &=&   {1\over 3}\, \bar{u} u
               +    {1\over 3}\, \bar{d} d \;, \\
  J^{5}(q)     &=&   {1\over 5}\, \bar{u} \gamma^{5} u
               -    {1\over 5}\, \bar{d} \gamma^{5} d \;.
 \end{eqnarray}
 \label{scurrent}
\end{mathletters}
The charges $g^{\rm s}_i=1/3$, $g^5_u=1/5$, and $g^5_d=-1/5$
have been chosen such that $g_N^{\sigma}=1$, $g_p^{\pi}=1$, and
$g_n^{\pi}=-1$ when $\Delta m=0$. Evaluating the nonrelativistic
reduction of Eqs.~(\ref{vertexsig}a) and (\ref{vertexsig}b)
in the quark model, we find
\begin{mathletters}
\label{thesigpi}
 \begin{eqnarray}
    g^{\sigma}_N  &=& \sum_{i=1}^3
    g^{{\rm s}}_i
    \langle N \!\uparrow | 1 | N\! \uparrow\rangle  \;, \\
   {g^{\pi}_N \over {2M_N}} &=& \sum_{i=1}^3
    \mu^{5}_i
    \langle N \!\uparrow | \sigma^z_i | N\! \uparrow\rangle   \;,
 \end{eqnarray}
\end{mathletters}
where we have defined $\mu^5_i\equiv g^5_i/2m_i$. From
Eq.~(\ref{thesigpi}a) we see that the sigma meson generates
merely a spin-independent coupling to the nucleon in the
nonrelativistic limit, so that there is no isospin breaking
in the $NN\sigma$ vertex and no contribution from sigma exchange
to the CSB potential. Thus, we will not consider sigma exchange
further. However, the quark mass dependence contained in $\mu^5_i$
in Eq.~(\ref{thesigpi}b) implies that the isospin-breaking in the
pion case is finite. The breaking to ${\cal O}(\Delta m/m)$ is
indicated in Table~\ref{tabletwo}. Note, however, that the computed
breaking at $q^2=0$ depends on the nature of the assumed pion-nucleon
coupling. If we had chosen {\it pseudovector} coupling, rather,
then no isospin breaking would result. The pseudovector current
contains no quark mass dependent pieces in the nonrelativistic limit.
Thus, our prediction in the $\pi^0$ case is decidedly more model
dependent than in the $\rho^0$ and $\omega$ channels. Moreover,
in the latter case, the compatibility of the computed tensor-to-vector
coupling constant ratios with the Bonn potential indicates that vector
dominance, which we assume, has some phenomenological support. Note that
in the $\pi^0$ case, there is no such independent support of our
``pseudoscalar dominance'' assumption. This concludes our discussion
of isospin breaking in the $NN$-meson vertices.

\section{Charge symmetry breaking potentials}
\label{seccsbp}

We shall now compute the CSB potentials which arise from the
isospin-violating couplings computed in the previous section
and tabulated in Table~\ref{tabletwo}. In an one-boson
exchange approximation, presuming the form of the
isospin breaking found in the $q^2=0$ results,
we obtain the following CSB potentials
for $\omega$, $\rho^0$, and $\pi^0$ exchange, respectively:
\begin{mathletters}
 \begin{eqnarray}
  \widehat{V}^{\omega}_{\rm CSB} &=& {V}^{\omega}_{\rm CSB}
   \Big[ \Gamma^{\mu}(1) \gamma_{\mu}(2) \tau_z(1) +
         \gamma^{\mu}(1) \Gamma_{\mu}(2) \tau_z(2) \Big]
   \;, \label{vomegaa} \\
  \widehat{V}^{\rho}_{\rm CSB} &=&  {V'}^{\rho}_{\rm CSB}
   \Gamma^{\mu}(1) \Gamma_{\mu}(2) \Big[ \tau_z(1) + \tau_z(2) \Big] +
 {V}^{\rho}_{\rm CSB}
 \Big[ \Gamma^{\mu}(1) \gamma_{\mu}(2) \tau_z(2) +
         \gamma^{\mu}(1) \Gamma_{\mu}(2) \tau_z(1) \Big]
   \;, \label{vrhoa} \\
  \widehat{V}^{\pi}_{\rm CSB} &=&  {V}^{\pi}_{\rm CSB}
   \gamma^{5}(1) \gamma^{5}(2) \Big[ \tau_z(1) + \tau_z(2) \Big]
   \;, \label{vpiona}
 \end{eqnarray}
\end{mathletters}
where $\Gamma^{\mu} \equiv i\sigma^{\mu\nu}(p'-p)_{\nu}/2M$
and we have defined
\begin{mathletters}
 \begin{eqnarray}
 {V}^{\omega}_{\rm CSB}(q) &\equiv&
   -\left({g_{\lower 2pt \hbox{$\scriptstyle NN\omega$}}^2
          \over q^2 - m_{\omega}^2}\right)f^\omega_1 g^\omega_0
    \;, \label{vomegab} \\
 {V}^{\rho}_{\rm CSB}(q) &\equiv&
   -\left({g_{\lower 2pt \hbox{$\scriptstyle NN\rho$}}^2
          \over q^2 - m_{\rho}^2}\right)f^\rho_0 g^\rho_1
    \;, \label{vrhob} \\
 {V'}^{\rho}_{\rm CSB}(q) &\equiv&
   -\left(
{g_{\lower 2pt \hbox{$\scriptstyle NN\rho$}}^2
          \over q^2 - m_{\rho}^2 }\right)f^\rho_0 f^\rho_1
\;, \label{vprhob} \\
 {V}^{\pi}_{\rm CSB}(q) &\equiv&
   -\left({g_{\lower 2pt \hbox{$\scriptstyle NN\pi$}}^2
          \over q^2 - m_{\pi}^2}\right)g^\pi_0 g^\pi_1
    \;. \label{vpionb}
 \end{eqnarray}
\end{mathletters}
Isospin breaking in the meson-nucleon vertices give rise to
the above CSB potentials, as per Eqs.~(\ref{vomegab})-(\ref{vpionb}).
The isospin-conserving tensor coupling is nonzero in the case of the
$\rho$ vertex, so that an additional potential of strength
${V'}_{\rm CSB}^{\rho}(q)$ arises.  These contributions have been
considered only recently~\cite{ghp95}. Yet the potentials of
Eqs.~(\ref{vomegaa}) and (\ref{vpiona}) are identical in form to
those generated by $\rho$-$\omega$ and $\pi$-$\eta$ mixing, respectively.
That is,
\begin{mathletters}
 \begin{eqnarray}
  \widehat{V}^{\rho\omega}_{\rm CSB} &=& {V}^{\rho\omega}_{\rm CSB}
   \Big[ \Gamma^{\mu}(1) \gamma_{\mu}(2) \tau_z(1) +
         \gamma^{\mu}(1) \Gamma_{\mu}(2) \tau_z(2) \Big]
   \;, \label{vrhoomegaa} \\
  \widehat{V}^{\pi\eta}_{\rm CSB} &=&  {V}^{\pi\eta}_{\rm CSB}
   \gamma^{5}(1) \gamma^{5}(2) \Big[ \tau_z(1) + \tau_z(2) \Big]
   \;, \label{vpietaa}
 \end{eqnarray}
\end{mathletters}
where
\begin{mathletters}
 \begin{eqnarray}
 {V}^{\rho\omega}_{\rm CSB}(q) &\equiv& -
  {f_{\lower 2pt \hbox{$\scriptstyle NN\rho$}}
   g_{\lower 2pt \hbox{$\scriptstyle NN\omega$}}
  \over (q^2 - m_{\rho}^2 )(q^2 - m_{\omega}^2 )}
  \langle \rho | H | \omega \rangle  \;,
 \label{vrhoomegab} \\
 {V}^{\pi\eta}_{\rm CSB}(q)    &\equiv& -
  {g_{\lower 2pt \hbox{$\scriptstyle NN\pi$}}
   g_{\lower 2pt \hbox{$\scriptstyle NN\eta$}}
   \over (q^2 - m_{\pi}^2 )(q^2 - m_{\eta}^2 )}
   \langle \pi | H | \eta \rangle  \;.
   \label{vpietab}
 \end{eqnarray}
\end{mathletters}
Note that in Eq.~(\ref{vrhoomegab}) we introduce
$f_{\lower 2pt \hbox{$\scriptstyle NN\rho$}}$, the
phenomenological tensor coupling of the Bonn model~\cite{machl89}.
Rather than performing a nonrelativistic reduction of the
potentials in Eqs.~(\ref{vomegaa})-(\ref{vpiona}) and
Eqs.~(\ref{vrhoomegaa})-(\ref{vpietaa}), we simply classify the
former as, either, ``$\rho\omega-\!\!$~like'' or ``$\pi\eta-\!\!$~like''
potentials. The effect of these new isospin-violating potentials on
CSB observables can then be readily elucidated. For example, the
contribution from omega-meson exchange is identical in structure to
that from $\rho$-$\omega$ mixing and thus contributes as well
to $\Delta A$ in elastic $\vec{n}-\vec{p}$ scattering. Indeed,
we now show that the contribution from omega-meson exchange is
comparable in magnitude and identical in sign to the one obtained
from $\rho$-$\omega$ mixing --- if the mixing amplitude is fixed at
its on-shell value.

\subsubsection{One-boson exchange potentials of the $\rho$-$\omega$ kind}
Potentials of the form given in Eq.~(\ref{vrhoomegaa}) give rise to
class III and class IV CSB potentials. They are generated by the
interference between the isospin-conserving vector coupling and the
isospin-violating tensor coupling; note, for example, Eq.~(\ref{vomegaa})
and the second term in Eq.~(\ref{vrhoa}). Unlike the case of the omega,
the isospin structure of rho exchange is not identical to that of
$\rho$-$\omega$ mixing; they are related by exchanging $\tau_{z}(1)
\leftrightarrow \tau_{z}(2)$. Thus, rho exchange contributes to the
class IV $\rho$-$\omega$ mixing potential with a sign opposite to that
of the omega. No sign changes are necessary when computing its
$\pi\eta-\!\!$~like or class III $\rho\omega-\!\!$~like contribution.
Note that the contribution from rho exchange is small relative to that
from the omega --- this emerges despite the larger isospin-violating
coupling associated with the rho vertex (see Table~\ref{tabletwo}).
The vector $NN\rho$ coupling is simply small relative to that of the
omega; in the Bonn potential
$g_{\lower 2pt \hbox{$\scriptstyle NN\omega$}}^2 /
 g_{\lower 2pt \hbox{$\scriptstyle NN\rho$}}^2 \approx 27$~\cite{machl89}.
The relative importance of the various contributions can be estimated
by computing the CSB potentials at $q^2=0$. Recall that in this limit
the isospin-violating couplings are insensitive to the quark momentum
distribution; they depend only on the spin-flavor symmetry of the wave
function. Using the Bonn B potential parameters of Table~\ref{tablethree}
and a value for the quark-mass difference of
$\Delta m=4.1$~MeV~\cite{licht89}, we obtain the following results
at $q^2=0$:
\begin{mathletters}
 \begin{eqnarray}
 {V}^{\omega}_{\rm CSB}(q^2=0) &=&
     {g_{\lower 2pt \hbox{$\scriptstyle NN\omega$}}^2
     \over m_{\omega}^2} f^\omega_1 g^\omega_0
     \approx 2.49~{\rm GeV}^{-2} \;, \\
 {V}^{\rho}_{\rm CSB}(q^2=0) &=&
     {g_{\lower 2pt \hbox{$\scriptstyle NN\rho$}}^2
     \over m_{\rho}^2} f^\rho_0 g^\rho_1
     \approx 0.18~{\rm GeV}^{-2} \;.  \\
 {V}^{\rho\omega}_{\rm CSB}(q^2=0)  &=& -
     {f_{\lower 2pt \hbox{$\scriptstyle NN\rho$}}
      g_{\lower 2pt \hbox{$\scriptstyle NN\omega$}}
     \over m_{\rho}^2 m_{\omega}^2}
     \langle \rho | H | \omega \rangle \Big|_{q^2=0}
     =0 \;,
 \end{eqnarray}
\end{mathletters}
Several remarks are in order. First, the $\rho$-$\omega$ mixing
amplitude, if modeled via fermion loops~\cite{piewil93,oconn94},
necessarily vanishes at $q^2=0$ in our model. Our model assumes
vector dominance, so that the vector-meson--nucleon vertices are
determined by the appropriate isospin components of the quark
electromagnetic current. Thus, the vector mesons couple to
currents that are conserved at the nucleon level, so that the
above result follows~\cite{piewil93,oconn94}. At the $q^2=0$
point, the charge-symmetry violation in our model comes
purely from the vertex contributions. Note that the rho meson
contribution to the latter is, indeed, small. It represents
merely a 7\% correction to the contribution from one-omega
exchange. Second, the strength of the CSB potentials generated
from omega exchange is comparable in magnitude to those obtained
from $\rho$-$\omega$ mixing if the {\it on-shell} value of the
mixing amplitude is assumed,
$\langle\rho|H|\omega\rangle|_{q^2=m_\omega^2}=-4520 \pm
600~{\rm MeV}^2$~\cite{coon87}.
Note, moreover, that the $\omega$ and on-shell $\rho$-$\omega$
mixing contributions are {\it identical} in sign. Specifically,
\begin{equation}
 {\widetilde V}^{\rho\omega}_{\rm CSB}(q^2=0)  = -
     {f_{\lower 2pt \hbox{$\scriptstyle NN\rho$}}
      g_{\lower 2pt \hbox{$\scriptstyle NN\omega$}}
     \over m_{\rho}^2 m_{\omega}^2}
     \langle \rho | H | \omega \rangle \Big|_{q^2=m_\omega^2}
     \approx 2.07~{\rm GeV}^{-2} \;,
\end{equation}
A CSB potential of this magnitude is needed for a successful
description of $\Delta A$ at 183 MeV~\cite{knut90}. Summing
our omega exchange contribution to the CSB potential and that
from on-shell $\rho$-$\omega$ mixing is not only internally
inconsistent but also gives a final potential which is too
large to fit the data --- see Sec.~\ref{secresults}. Our results
suggest that a class IV potential of the appropriate size is
generated by isospin-violations in the $NN\omega$ vertex, together
with small corrections from rho exchange and off-shell
$\rho$-$\omega$ mixing. This is the central result of our paper.

\subsubsection{One-boson exchange potentials of the $\pi$-$\eta$ kind}
Potentials of the form given in Eq.~(\ref{vpietaa}) generate class III
CSB potentials exclusively. The Lorentz structure of the first term in
Eq.~(\ref{vrhoa}) differs from that of the $\pi$-$\eta$ mixing and
one-pion exchange potentials, so that it is convenient to perform a
nonrelativistic reduction of all three contributions, i.e.,
\begin{mathletters}
 \begin{eqnarray}
  \widehat{V}^{\pi\eta}_{\rm CSB} &=& - {V}^{\pi\eta}_{\rm CSB}(q)
    \left( {{\bf q}^2 \over 12M^2} \right)
    \Big[ \mbox{\boldmath$\sigma$}_1 \cdot \mbox{\boldmath$\sigma$}_2 +
           S_{12}(\hat{\bf q}) \Big]
    \Big[ \tau_z(1) + \tau_z(2) \Big] \;, \label{vpietac}  \\
  \widehat{V}^{\pi}_{\rm CSB} &=& - {V}^{\pi}_{\rm CSB}(q)
    \left( {{\bf q}^2 \over 12M^2} \right)
    \Big[ \mbox{\boldmath$\sigma$}_1 \cdot \mbox{\boldmath$\sigma$}_2 +
          S_{12}(\hat{\bf q}) \Big]
    \Big[ \tau_z(1) + \tau_z(2) \Big] \;, \label{vpionc}  \\
  \widehat{V}^{\prime\rho}_{\rm CSB} &=& - {V}^{\prime\rho}_{\rm CSB}(q)
    \left( {{\bf q}^2 \over 12M^2} \right)
    \Big[ 2\mbox{\boldmath$\sigma$}_1 \cdot \mbox{\boldmath$\sigma$}_2 -
          S_{12}(\hat{\bf q}) \Big]
    \Big[ \tau_z(1) + \tau_z(2) \Big] \;, \label{vrhoc}
 \end{eqnarray}
\end{mathletters}
where we have introduced the tensor operator $S_{12}(\hat{\bf q}) =
[3(\mbox{\boldmath$\sigma$}_1\cdot{\hat{\bf q}})
  (\mbox{\boldmath$\sigma$}_2\cdot{\hat{\bf q}}) -
   \mbox{\boldmath$\sigma$}_1\cdot\mbox{\boldmath$\sigma$}_2]$.
We estimate the relative size of these contributions by evaluating them
at $q^2=0$, noting table~\ref{tablethree}:
\begin{mathletters}
\label{zeroq2pieta}
\begin{eqnarray}
 {V}^{\pi}_{\rm CSB}(q^2=0) &=&
     {g_{\lower 2pt \hbox{$\scriptstyle NN\pi$}}^2
     \over m_{\pi}^2} g^\pi_0 g^\pi_1
     \approx 36.83~{\rm GeV}^{-2} \;, \\
 {V}^{\prime\rho}_{\rm CSB}(q^2=0) &=&
     {g_{\lower 2pt \hbox{$\scriptstyle NN\rho$}}
      f_{\lower 2pt \hbox{$\scriptstyle NN\rho$}}
     \over m_{\rho}^2} f^\rho_0
     \approx 1.07~{\rm GeV}^{-2} \;,
 \end{eqnarray}
\end{mathletters}
%
The one-pion exchange contribution dominates that of the rho;
this is driven by the large $\rho$-$\pi$ mass difference --- recall
$m_{\rho}^{2}/m_{\pi}^{2}\approx 30$. Note that the inclusion
of the rho meson leads to a reduction of the tensor and an
enhancement of the spin-spin components of the pion-exchange
potential. Unlike the vector meson case, we cannot readily
compute the $\pi$-$\eta$ mixing amplitude at $q^2=0$ in our model.
That is, in the pion case, there is no conserved current, so that
the $q^2=0$ mixing can be nonzero~\cite{piekar93,maltman93,chan95}.
Nevertheless, we can compare the results of Eq.~(\ref{zeroq2pieta})
with the ``usual'' $\pi$-$\eta$ mixing potential:
\begin{equation}
 {V}^{\pi\eta}_{\rm CSB}(q^2=0) = -
     {g_{\lower 2pt \hbox{$\scriptstyle NN\pi$}}
      g_{\lower 2pt \hbox{$\scriptstyle NN\eta$}}
     \over m_{\pi}^2 m_{\eta}^2}
     \langle \pi | H | \eta \rangle \Big|_{q^2=m_\eta^2}
     \approx 52.01~{\rm GeV}^{-2} \;,
\label{pietamix}
\end{equation}
where we have input the $\pi$-$\eta$ mixing matrix element evaluated
at its on-shell point, $\langle\pi|H|\eta\rangle|_{q^2=m_\eta^2}=
-4200$~MeV$^{2}$~\cite{coon82}. The contribution from one-pion exchange
is comparable to that from $\pi$-$\eta$ mixing. The $\pi$-$\eta$ mixing
potential may seem slightly larger, but the $NN\eta$ coupling is
ill-determined from two-nucleon data. Indeed, it is believed that the
Bonn potential overestimates it --- a current analysis based on
$\eta$-photoproduction data suggests couplings as low as
$g^2_{NN\eta}/4\pi \alt 0.5$~\cite{tiator94} (see also
Ref~\cite{piekar93}).

The CSB potentials from one-pion exchange have been computed previously
in a nucleon model~\cite{cheung80}. Here the neutron-proton mass
difference, $\Delta M$, generates the breaking. In the specific case
of the class III contribution coming from neutral pion exchange, the
scale of the breaking is set by $\Delta M/2M$. Thus, the isospin breaking
in the quark model is substantially larger than in the nucleon model,
i.e.,
\begin{equation}
   \left({3 \over 10}{\Delta m \over m}\right) \bigg/
   \left({1 \over 2} {\Delta M \over M}\right) \approx 6 \;,
\end{equation}
so that any CSB observable receiving an important contribution
from $\pi$-$\eta$ mixing will also be affected by the exchange
of neutral pions. The breaking we calculate in the $NN\pi^0$ vertex
is identical to the result of Mitra and Ross~\cite{miller90,mitra67}.
Note that the exchange of charged pions --- and rhos ---
generates a class IV potential which is important in the analysis
of $\Delta A$~\cite{willia87,miller90}. In the charged meson
case, however, the relation between the isospin-violating couplings
in the two models is not simple: it depends on the quark momentum
distribution. Yet, under reasonable assumptions, both sorts of models
seem to generate class IV potentials of comparable
strength~\cite{miller90}.

\section{Results}
\label{secresults}

In this section we compute the CSB potentials for a range of
spacelike momenta. We shall concentrate on class IV contributions
exclusively as we are interested in computing the impact of
the new isospin-violating sources on $\Delta A$. The knowledge of
the $q^2=0$ couplings now no longer suffices. One is forced to model
the momentum dependence of the coupling constants --- including that
of the isospin-violating components. Here we consider two different
estimates for the $q^2$ dependence of the CSB potentials. First, we
simply adopt the momentum dependence which emerges from fits to the
isospin-conserving two-nucleon data. Thus, the ratio of the
isospin-violating to the isospin-conserving coupling, e.g.,
$f^{\omega}_{1}/g^{\omega}_{0}$, remains unchanged. Note that
in the Bonn model $f^{\rho}_{1}/g^{\rho}_{1}$ is also a constant.
We implement this choice by modifying the meson-nucleon ``point''
couplings indicated in Eq.~(\ref{vert}) as per the Bonn B potential
parameters, see Table~\ref{tablethree}. That is,
\begin{mathletters}
\label{bonnff}
 \begin{eqnarray}
    g_{\lower 2pt \hbox{$\scriptstyle NN\omega$}} \rightarrow
    g_{\lower 2pt \hbox{$\scriptstyle NN\omega$}}({\bf q}^2) &=&
    g_{\lower 2pt \hbox{$\scriptstyle NN\omega$}}
    (1+{\bf q}^2/\Lambda^2_{\omega})^{-2} \;, \\
    g_{\lower 2pt \hbox{$\scriptstyle NN\rho$}} \rightarrow
    g_{\lower 2pt \hbox{$\scriptstyle NN\rho$}}({\bf q}^2) &=&
    g_{\lower 2pt \hbox{$\scriptstyle NN\rho$}}
    (1+{\bf q}^2/\Lambda^2_{\rho})^{-2} \;.
 \end{eqnarray}
\end{mathletters}
This is an additional model assumption.
Here we use {\bf q} to denote the three-momentum transfer; we consider
the form factors in the Breit frame, where $q_0=0$ and $q^2=-{\bf q}^2$.
Second, we compute the
${\cal O}({\bf q}^2)$ isospin-breaking in the couplings in the
nonrelativistic quark model, in order to gauge the uncertainty
in the momentum dependence of the CSB potentials. Let us examine
the isospin-breaking in the Sachs-Walecka form factors~\cite{sachswal},
separated into contributions from the isoscalar or isovector quark
charges. These quantities are related to the $\omega$ and $\rho$
couplings by virtue of our vector dominance assumption. As previously,
we will discuss merely the isospin breaking in the $NN\omega$ vertex in
detail. Now
\begin{mathletters}
 \begin{eqnarray}
    G_{E,p}^{\omega} &=&
      2 g^{+}_u \langle u \rangle_p + g^{+}_d \langle d \rangle_p \;, \\
    G_{E,n}^{\omega} &=&
      2 g^{+}_d \langle d \rangle_n + g^{+}_u \langle u \rangle_n \;,
 \end{eqnarray}
 \label{gepm}
\end{mathletters}
and
\begin{mathletters}
 \begin{eqnarray}
 {G_{M,p}^{\omega} \over 2 M_p} &=&
    {1 \over 18}\left( {4\over m_u} \langle u \rangle_p
       - {1\over m_d} \langle d \rangle_p \right)
\;, \\
 {G_{M,n}^{\omega} \over 2 M_n} &=&
    {1 \over 18}\left( {4\over m_d} \langle d \rangle_n
       - {1\over m_u} \langle u \rangle_n \right)
\; .
 \end{eqnarray}
 \label{gmpm}
\end{mathletters}
These expressions are generalizations of Eqs.~(\ref{model1}) and
(\ref{model2}). We have used the notation of Eq.~(\ref{themodel})
in denoting the isoscalar and isovector quark charges and have
introduced $\langle u \rangle_p$, for example, to represent the
Fourier transform of the proton wave function with respect to the
up quark coordinate. We compute the latter in the harmonic
oscillator quark model for simplicity. In the harmonic oscillator
quark model~\cite{dmitra95,bhaduri} the nucleon possesses a mass
$M_N = 2m_1 + m_2$, so that for the proton $m_1=m_u$ and $m_2=m_d$.
For convenience one defines $R^{-2}_{\rho}=\sqrt{3k m_{\rho}}$ and
$R^{-2}_{\lambda}=\sqrt{3k m_{\lambda}}$, where
$m_{\lambda} = 3m_1m_2 / (2m_1 + m_2)$, $m_{\rho} = m_1$, and
$k$ is the spring constant. One finds that~\cite{dmitra95}
\begin{mathletters}
  \begin{eqnarray}
  \langle u \rangle_p &\equiv& \langle
  \exp({\it i} {\bf q}\cdot{\bf r}_u) \rangle_p = 1 - {{\bf q}^2 \over 8}
  \left( R_{\rho p}^2 + 3\left(m_d \over M_p\right)^2 R_{\lambda p}^2 \right)
  + {\cal O}({\bf q}^4) \;, \\
  \langle d \rangle_p &\equiv& \langle
  \exp({\it i} {\bf q}\cdot{\bf r}_d) \rangle_p = 1 - {3{\bf q}^2 \over 2}
  \left(m_u \over M_p\right)^2 R_{\lambda p}^2 + {\cal O}({\bf q}^4) \;, \\
  \langle u \rangle_n &\equiv& \langle
  \exp({\it i} {\bf q}\cdot{\bf r}_u) \rangle_n = 1 - {3{\bf q}^2 \over 2}
  \left(m_d \over M_n\right)^2 R_{\lambda n}^2 + {\cal O}({\bf q}^4) \;, \\
  \langle d \rangle_n &\equiv& \langle
  \exp({\it i} {\bf q}\cdot{\bf r}_d) \rangle_n = 1 - {{\bf q}^2 \over 8}
  \left( R_{\rho n}^2 + 3\left(m_u \over M_n\right)^2 R_{\lambda n}^2 \right)
  + {\cal O}({\bf q}^4) \;.
  \end{eqnarray}
\label{qformf}
\end{mathletters}
We write the Fourier transforms in Eq.~(\ref{qformf}) through
${\cal O}({\bf q}^2)$ only. This suffices to make contact with
the hadronic form factors. Moreover, one cannot expect the
nonrelativistic quark model to be reliable at still larger momentum
transfers. We must now relate the above electric and magnetic form
factors to the $f$'s and $g$'s present in the definition of the vertex,
Eq.~(\ref{vert}). Following the usual relation between the electromagnetic
form factors $G_E$, $G_M$ and $F_1$, $F_2$, vector dominance dictates that
\begin{mathletters}
\label{vectdom}
\begin{eqnarray}
 G_{E,N}^{\omega}(q^2) &=& g_N^{\omega}(q^2)
+ {q^2 \over 4 M_N^2} f_N^{\omega}(q^2) \;,
\\
 G_{M,N}^{\omega}(q^2) &=& g_N^{\omega}(q^2) + f_N^{\omega}(q^2)
\;.
\end{eqnarray}
\end{mathletters}
Fits to the electromagnetic form factor data indicate that
$F_1(q^2)$ and $F_2(q^2)$ fall with different rates in $q^2$;
vector dominance implies that this should be true of
$g_N^{\omega,\rho}$ and $f_N^{\omega,\rho}$ as well. Note, this
is at odds with the Bonn model, as it assumes that the ratio
$f_N^{\omega,\rho}/g_N^{\omega,\rho}$ is constant.
We proceed as follows. We compute $f_N^{\omega}$ and $g_N^{\omega}$
to ${\cal O}({\bf q}^4,{\Delta m}^2)$, using
Eqs.~(\ref{gepm}-\ref{vectdom}). Then we estimate the ``effective''
$\Lambda_{\omega}$, as defined in Eq.~(\ref{bonnff}), required to
reproduce the isospin breaking computed to
${\cal O}({\bf q}^4,{\Delta m}^2)$ and use that
$\Lambda_{\omega}$ in our subsequent computation of
the spin-singlet--triplet mixing angles. Thus,
\begin{mathletters}
\begin{eqnarray}
 g_{0}^{\omega} + g_{1}^{\omega}\tau_z &=& \left[
  1 - {{\bf q}^2 R^2 \over 6} \right] +
  {\Delta m \over m} \left[ {5{\bf q}^2 \over 24 M^2} -
  {{\bf q}^2 R^2 \over 72} \right] \tau_z             +
  {\cal O}({\bf q}^4,{\Delta m}^2) \;, \\
 f_{0}^{\omega} + f_{1}^{\omega}\tau_z &=&
 {5\over 6}{\Delta m \over m}
 \left[ 1 - {{\bf q}^2 R^2 \over 3} -
 {{\bf q}^2 \over 4 M^2} \right] \tau_z +
 {\cal O}({\bf q}^4,{\Delta m}^2) \;.
\label{breako}
\end{eqnarray}
\end{mathletters}
%
Several remarks are in order. First, note that we have defined
$R^{-2}=\sqrt{3k m}$, where $m$ is the average mass of the up
and down quarks. From Eq.~(\ref{breako}) we observe
$f_0^{\omega}=0$ to ${\cal O}({\bf q}^4,\Delta m^2)$; this is
consistent with the Bonn model, which assumes $f_0^{\omega}=0$
for all $q^2$. We have performed the same calculations for
the $NN\rho$ vertex as well. In this case, one finds results at
odds with the Bonn model, as $f_1^{\rho}/g_1^{\rho}$ is not constant
to ${\cal O}({\bf q}^4,\Delta m^2)$. Note that at nonzero ${\bf q}^2$
CSB potentials beyond those enumerated in
Eqs.~(\ref{vomegab})-(\ref{vpionb}) may exist. For example, at
${\cal O}({\bf q}^2,\Delta m)$ a new CSB contribution arises
from the combination $f_1^{\rho}g_0^{\rho}$. Yet, like the rho
contribution to the CSB potential given in Eq.~(\ref{vrhob}), it
is not numerically important, due to the small value of
$g_{\lower 2pt \hbox{$\scriptstyle NN\rho$}}$ in the Bonn
model --- recall that $g_{\lower 2pt \hbox{$\scriptstyle NN\omega$}}^2 /
 g_{\lower 2pt \hbox{$\scriptstyle NN\rho$}}^2 \approx 27$~\cite{machl89}.
Let us proceed to examine the impact of Eq.~(\ref{breako}) on the omega
contribution to the class IV CSB potential. We fix the scale $R$ by
requiring that the isospin conserving vertex, $g_0^{\omega}$, fall in
${\bf q}^2$ at the rate given by the Bonn model, so that
$R=\sqrt{12}/\Lambda_{\omega} \approx .37$~fm. We choose $R$ in this
manner as our primary interest is in determining the fall-off of the
isospin-breaking potential {\it relative} to the isospin-conserving one.
Noting Eq.~(\ref{vomegab}), we consider
\begin{mathletters}
\begin{eqnarray}
f_{1}^{\omega} g_{0}^{\omega} &=&
 {5 \over 6} {\Delta m \over m}
 \left( 1 - 6{{\bf q}^2 \over \Lambda_{\omega}^2} -
 {{\bf q}^2 \over 4 M^2} \right) +
 {\cal O}({\bf q}^4,{\Delta m}^2) \\
 \quad &\equiv& {5 \over 6} {\Delta m \over m}
 \left( 1 - 4{{\bf q}^2 \over {\widetilde\Lambda}_{\omega}^2} +
 {\cal O}({\bf q}^4) \right) \;.
\end{eqnarray}
\end{mathletters}
By replacing the $\Lambda_{\omega}$ of Eq.~(\ref{bonnff}) with
the ${\widetilde\Lambda}_{\omega}$ given above, such that
\begin{equation}
{\widetilde\Lambda}_{\omega}^2 = \Lambda_{\omega}^2
\left( { 4 \over 6 + \Lambda_{\omega}^2 /4 M^2 } \right) \;,
\label{tildel}
\end{equation}
we obtain an expression for the CSB potential, Eq.~(\ref{vomegab}),
which is of the form given by our original prescription, Eq.~(\ref{bonnff}),
yet is equivalent to the isospin-breaking calculated in the harmonic
oscillator quark model at ${\cal O}({\bf q}^2,\Delta m)$. Numerically,
the Bonn model $\Lambda_{\omega}=1850$ MeV is changed to
$\widetilde{\Lambda}_{\omega}=1401$ MeV. At this order
the coefficient of $g_1^{\rho}$ is not negative, so that we cannot carry
out the above exercise for the rho as well. The rho's numerical impact
on the CSB potential is small, so that this gap does not impact our
uncertainty estimate in any significant way. We will proceed to compute
the spin-singlet--triplet mixing angles for the potential given by
Eqs.~(\ref{vomegab}) and (\ref{bonnff}) for both the $\Lambda_{\omega}$
of the Bonn potential and the $\widetilde{\Lambda}_{\omega}$ of
Eq.~(\ref{tildel}).

In Fig.~\ref{figone} we present estimates of the CSB potentials given
in Eqs.~(\ref{vomegab}), (\ref{vrhob}), and (\ref{vrhoomegab}) using
Eq.~(\ref{bonnff}) with the $\Lambda_{\omega,\rho}$ of the Bonn model.
The qualitative conclusions we draw here are not sensitive to the choice
of $\Lambda_{\omega,\rho}$, so that we simply present the potentials
computed in the Bonn model. The solid line is the CSB potential which
results from $\rho$-$\omega$ mixing, Eq.~(\ref{vrhoomegab}), if
the on-shell value of the $\rho$-$\omega$ mixing amplitude is employed
for the entire range of momenta. This is the potential traditionally
used in studies of CSB observables. A potential of this strength is
required to describe the analyzing power difference  $\Delta A$
measured in elastic $\vec{n}-\vec{p}$ scattering~\cite{holz87}. In
contrast, the dashed line results if the momentum-dependent
$\rho$-$\omega$ mixing amplitude of Ref.~\cite{piewil93} is employed
in Eq.~(\ref{vrhoomegab}) --- this is too small to
fit the data~\cite{iqnisk94}, yet
a model in which the vector mesons couple to conserved currents
must yield a vanishing mixing amplitude at $q^2=0$~\cite{oconn94}.
We have not extended our model to describe $\rho$-$\omega$ mixing;
the vector dominance assumption we use implies, however, that the
$q^2=0$ mixing must be zero in this framework. We take the
momentum-dependence of the mixing amplitude computed by
Piekarewicz and Williams~\cite{piewil93} as archetypal. This latter
CSB potential in itself would upset the previous agreement with
experiment. However, the new sources of isospin violation computed
here are sufficient to restore the agreement. In particular, the
contribution from omega-meson exchange, given by the dash-dotted line,
is large and comparable in magnitude to the one arising from on-shell
$\rho$-$\omega$ mixing. We have also computed the contribution from
the rho-meson, given by the dotted line, though it is negligible due
to the small $NN\rho$ vector coupling.

In Fig.~\ref{figtwo} we display the above CSB potentials in
configuration space. The potentials have been normalized so
that the areas under the curves equal $V(q^2=0)$. Qualitatively,
the trends observed in Fig.~\ref{figone} remain: we obtain large
contributions from on-shell $\rho$-$\omega$ mixing and omega-meson
exchange and small corrections to the latter from off-shell
$\rho$-$\omega$ mixing and rho-meson exchange. These results are
suggestive, yet we can obtain a precise estimate of the impact of
the enumerated isospin-violating sources on $\Delta A$ by calculating
the spin-singlet--triplet mixing angles,
$\bar{\gamma}_{\lower 2pt \hbox{$\scriptstyle J$}}$.
These are the dynamical quantities driving
$\Delta A$~\cite{willia87,miller90}.
Recall that the elastic scattering amplitude of two spin-$1/2$
particles is specified by six invariant amplitudes
$a, b, c, d, e,$ and $f$~\cite{miller90}, so that
\begin{eqnarray}
\widehat{M}
&=&{1\over 2} \Big[ (a + b) +
(a-b) ({\bf \sigma}_1\cdot \hat{\bf n}) ({\bf \sigma}_2\cdot \hat{\bf n})
\nonumber \\
&+& (c+d) ({\bf \sigma}_1\cdot \hat{\bf m}) ({\bf \sigma}_2\cdot \hat{\bf m})
+ (c-d) ({\bf \sigma}_1\cdot \hat{\bf l}) ({\bf \sigma}_2\cdot \hat{\bf l}) \\
&+& e ({\bf \sigma}_1 + {\bf \sigma}_2)\cdot \hat{\bf n}
+ f ({\bf \sigma}_1 - {\bf \sigma}_2)\cdot \hat{\bf n}
 \Big]
\nonumber \;,
\end{eqnarray}
where
\begin{equation}
\hat{\bf l}\equiv { {\bf k}_f + {\bf k}_i \over
{| {\bf k}_f + {\bf k}_i |}} \;, \quad
\hat{\bf m}\equiv { {\bf k}_f - {\bf k}_i \over
{| {\bf k}_f - {\bf k}_i |}} \;, \quad
\hat{\bf n}\equiv { {\bf k}_i \times {\bf k}_f \over
{| {\bf k}_i \times {\bf k}_f |}} \;,
\end{equation}
and ${\bf k}_i$ and ${\bf k}_f$ are the initial and final
c.m. momenta of particle 1. The $\vec{n}-\vec{p}$ analyzing power difference
is nonzero only if accompanied by spin-singlet--triplet mixing,
specifically
\begin{equation}
\Delta A(\theta) \equiv A_n(\theta) - A_p(\theta) =
2 {\rm Re } (b^* f)/\sigma_0 \;,
\end{equation}
where $\sigma_0$ is the unpolarized differential cross section.
The spin-singlet--triplet mixing is controlled by $f$. Neglecting
electromagnetic effects, $f$ is connected to the mixing angles
$\bar{\gamma}_{\lower 2pt \hbox{$\scriptstyle J$}}$ via~\cite{gersten81}
\begin{equation}
f(k,\theta) ={ i \over 2k} \sum_{J=1}^{\infty}
(2J +1) \sin (2   \bar{\gamma}_{\lower 2pt \hbox{$\scriptstyle J$}})
\exp ( i    \bar{\delta}_{\lower 2pt \hbox{$\scriptstyle J$}} +
i    \bar{\delta}_{\lower 2pt \hbox{$\scriptstyle JJ$}} )
d_{10}^J (\theta) \;,
\end{equation}
where the $d_{10}^J(\theta)$ are Wigner functions and the
$ \bar{\delta}_{\lower 2pt \hbox{$\scriptstyle J$}}$ and
$ \bar{\delta}_{\lower 2pt \hbox{$\scriptstyle JJ$}}$
are the singlet
and uncoupled triplet bar phase shifts, respectively.
In a distorted-wave Born approximation the mixing angles
themselves are given by~\cite{willia87}
\begin{equation}
   \bar{\gamma}_{\lower 2pt \hbox{$\scriptstyle J$}}
   = -4Mk \sqrt{J(J+1)}
   \int_0^\infty dr r^{2} R_{J}(r) V_{IV}(r) R_{JJ}(r)
 \equiv \int_0^\infty dr
 I_{{\lower 2pt \hbox{$\scriptstyle J$}}}(r)
 \label{mixangle} \;,
\end{equation}
where we have introduced the class IV CSB potential
\begin{equation}
  V_{IV}(r) \equiv {1 \over 2M^{2}r}
            {dV_{\rm CSB}(r) \over dr} \;.
\end{equation}
Note that
$R_{J}(r)$ and $R_{JJ}(r)$ are the spin-singlet and triplet
radial wave functions, respectively, for $NN$ scattering
in the $L=J$ channel. Distortion effects are incorporated
through these radial wave functions; we assume them adequately
described by solutions to the Reid soft-core
potential~\cite{reid68}. In Table~\ref{tablefour} the first
four nonvanishing mixing angles, $J=1 - 4$, are presented at a
laboratory energy of 183 MeV. In  addition, the integrand from
which $\bar{\gamma}_{1}$ is obtained, that is, $I_{1}(r) $
in Eq.~(\ref{mixangle}), is plotted in Fig.~\ref{figthree}.
This represents the class IV potential suitably weighted by
realistic $NN$ wave functions. Three calculations are presented
for comparison. The solid line is obtained using Eq.~(\ref{vrhoomegab})
and the on-shell value of the $\rho$-$\omega$ mixing amplitude; the
area under this curve is the mixing angle required to reproduce
the $\Delta A$ data. In the dashed line we have combined the
off-shell $\rho$-$\omega$ mixing contribution described above with
the isospin-violating vertex contributions arising from omega- and
rho-meson exchange. Albeit form factor uncertainties in the
isospin-violating vertices, this is our best estimate of the
mixing angle contribution. The vertices in this figure were
evaluated using Eq.~(\ref{bonnff}) and the Bonn cutoff parameters
$\Lambda_{\omega,\rho}$ tabulated in Table~\ref{tablethree}.
We have also combined the {\it on-shell} $\rho$-$\omega$ mixing
contribution with the above vertex contributions, even if our model
is not consistent with a nonzero mixing amplitude at $q^2=0$ --- this
is shown by the dash-dotted line. The integrand in this case is
considerably larger than the other two estimates. The $J=1$ mixing
angles for these integrands are displayed in parentheses next to the
curve labels. The agreement between the first two calculations
is very good. Indeed, the contribution from omega-meson exchange,
together with small corrections from off-shell $\rho$-$\omega$ mixing
and rho-meson exchange, results in a 3\% reduction in the value of
$\bar{\gamma}_{1}$, relative to the on-shell value. This kind of
agreement --- at the few percent level --- is maintained throughout
all the examined partial waves, note Table~\ref{tablefour}. These
computations have also been performed with the form factor
$\widetilde\Lambda_{\omega}$, Eq.~(\ref{tildel}), estimated in the
harmonic oscillator quark model. The mixing angles obtained in this
fashion vary by about 10\% in the important partial waves from those
computed with the Bonn form factors; note that
$\bar{\gamma}_{1}=.036^{\circ}$, rather than $.033^{\circ}$. For
a detailed comparison, see Table~\ref{tablefour} --- the mixing angles
which use the harmonic oscillator quark model results to estimate the
${\bf q}^2$ dependence of the CSB potentials are shown in parentheses.
The mixing angles computed with isospin-breaking meson-nucleon vertices
and off-shell $\rho$-$\omega$ mixing in the two approaches bracket the
old on-shell $\rho$-$\omega$ mixing results for $J=1-3$, so that these
new estimates are also quite close to the ``old'' on-shell results.
$\widetilde \Lambda_{\omega}$ is some 3/4 of $\Lambda_{\omega}$, yet
the above calculations show that the $\Delta A$ at 183 MeV is essentially
dominated by the $q^2=0$ physics. Note that if one were to assume a
momentum-independent $\rho$-$\omega$ mixing amplitude~\cite{miller94} and
to include the contributions from omega and rho exchange an increase of
almost a factor of two relative to the above mixing angle estimates would
result.

\section{Conclusions}
\label{secconcl}

 We have studied the charge-symmetry breaking in the $NN$
potential arising from isospin-violating meson-nucleon coupling
constants. The isospin-violating couplings are obtained by
computing matrix elements of quark currents of the appropriate
Lorentz and flavor structure between nucleon states. We have used
a nonrelativistic quark model to evaluate these matrix elements,
yet our estimates at $q^2=0$ depend merely on the spin and flavor
structure of the nucleon wave function, rather than on the details
of the quark momentum distribution. Thus, in the vector meson sector,
for example, our model estimates at $q^2=0$ depend on our vector
dominance assumption, but little else. We have also studied isospin
breaking in the $NN\pi$ and $NN\sigma$ vertices. No isospin-violations
exist in the $\sigma$ vertex at $q^2=0$. We have found that the breaking
in the $NN\pi$ vertex depends on whether the $\pi$N coupling is of
pseudoscalar or pseudovector character --- no isospin breaking results
if pseudovector coupling is assumed. However, a pseudoscalar $\pi$N
coupling is commonly used in studies of CSB, and the breaking we find
in the vertex is substantially larger than the breaking computed in
hadronic models of neutral pion exchange. Thus, any CSB observable
receiving an important contribution from $\pi$-$\eta$ mixing will also
be affected by the exchange of neutral pions.

 We have found that omega-meson exchange is an important component of the
class IV charge-symmetry-breaking $NN$ potential needed to describe the
analyzing power difference measured in elastic $\vec{n}-\vec{p}$ scattering
at low energies. The potential which emerges from the isospin-violating
$NN\omega$ vertex is identical in structure to that from $\rho$-$\omega$
mixing~\cite{ghp95}. Moreover, our estimates indicate that these two
contributions --- with the mixing amplitude fixed at its on-shell value
--- are comparable in magnitude and identical in sign at $q^2=0$. Models
in which the vector mesons couple to conserved currents, of which ours
is an example, have no $\rho$-$\omega$ mixing at $q^2=0$~\cite{oconn94}.
We have found that isospin-violation in the $NN\omega$ vertex can generate
a CSB potential of sufficient magnitude to fill the phenomenological role
required by the IUCF measurement of $\Delta A$ at 183 MeV.

 The isospin-violating couplings we have computed at $q^2=0$ do not suffice
to make a quantitative prediction of the CSB potential needed for the IUCF
experiment. One must compute the $q^2$ dependence of the $NN$~meson vertex
as well --- including that of the isospin-violating pieces. We have
considered two simple estimates. The first is simply a prescription:
we modify the ``point'' couplings by introducing hadronic form
factors according to the Bonn B potential. This assumes that the
relative strength of the isospin-breaking potential found at $q^2=0$
persists at finite $q^2$ as well. In the second we compute the isospin
breaking to ${\cal O}({\bf q}^4,\Delta m^2)$ using the spatial wave
functions of the harmonic oscillator quark model and find the hadronic
form factor for omega exchange needed to reproduce the isospin breaking
computed to the above order. The use of the spatial component of the
nucleon wave function is required here, so that this estimate is rather
more model dependent than our $q^2=0$ results. We find that the use of
the latter estimate yields slightly larger CSB potentials.

Armed with estimates of the momentum dependence of the $NN$~meson
vertex, we have computed the spin-singlet--triplet mixing  angles
$\bar{\gamma}_{\lower 2pt \hbox{$\scriptstyle J$}}$: these are the
fundamental dynamical quantities driving $\Delta A$.
Our $\bar{\gamma}_{\lower 2pt \hbox{$\scriptstyle J$}}$ computation
is realistic as we have used the Reid soft-core potential to generate
the distortions in the $NN$ wave function. We have computed the
spin-singlet--triplet angles using three different sources of isospin
violation: (1) $\rho$-$\omega$ mixing with the amplitude fixed at its
on-shell value, (2) off-shell $\rho$-$\omega$ mixing plus omega- and
rho-meson exchange, and (3) on-shell $\rho$-$\omega$ mixing plus omega-
and rho-meson exchange. The first case, used in the original estimates
of $\Delta A$, represents a ``baseline'' value, as it fits the data.
A CSB potential of this magnitude accounts for a substantial fraction
of the measured value of $\Delta A$ at 183 MeV. The second case, which
should be regarded as our best estimate, yields values for
$\bar{\gamma}_{\lower 2pt \hbox{$\scriptstyle J$}}$ that are
within 10\% of those obtained with on-shell $\rho$-$\omega$ mixing,
for the important partial waves. In contrast, case (3) results in a
factor-of-two enhancement relative to the original calculation using
on-shell $\rho$-$\omega$ mixing. Two important results thus emerge from
the present work. First, we have found a new source of isospin violation,
namely in the $NN\omega$ vertex, which can fill the role demanded by the
data. Second, we have shown that insisting upon a $\rho$-$\omega$ mixing
amplitude held constant at its on-shell value, after including the
contribution from omega-meson exchange, results in a class IV potential
too large to be consistent with the IUCF $\Delta A$ measurement.

\acknowledgments
We thank V. Dmitra\v{s}inovi\'c and S.J. Pollock for fruitful
discussions, A. Thomas for a helpful
suggestion, and S. Capstick for useful
conversations.
This work was supported by the DOE under
Contracts Nos. DE-FG02-87ER40365 (S.G. and C.J.H.),
DE-FC05-85ER250000 (J.P.), and DE-FG05-92ER40750 (J.P.).

\begin{figure}
 \caption{Charge-symmetry-breaking component of the $NN$ potential
          as a function of $q^{2}$ arising from
          on-shell  $\rho$-$\omega$ mixing (solid line),
          off-shell $\rho$-$\omega$ mixing (dashed line),
          omega-meson exchange (dash-dotted line), and
          rho-meson exchange (dotted line).}
 \label{figone}
\end{figure}

\begin{figure}
 \caption{Charge-symmetry-breaking component of the $NN$ potential
          as a function of $r$ arising from
          on-shell  $\rho$-$\omega$ mixing (solid line),
          off-shell $\rho$-$\omega$ mixing (dashed line),
          omega-meson exchange (dash-dotted line), and
          rho-meson exchange (dotted line).}
 \label{figtwo}
\end{figure}

\begin{figure}
 \caption{The integrand of the spin-singlet--triplet mixing angle
          $\bar{\gamma}_{1}$ at 183 MeV for three different
          estimates of CSB: on-shell $\rho$-$\omega$ mixing
          (solid line), off-shell $\rho$-$\omega$ mixing plus
          omega- and rho-meson exchange (dashed line), and
          on-shell $\rho$-$\omega$ mixing plus omega-
          and rho-meson exchange (dash-dotted line). The
          value of $\bar{\gamma}_{1}$ for each estimate, which is
          simply the area under the appropriate
          curve, appears in parentheses next to its
          label.}
 \label{figthree}
\end{figure}

\mediumtext
 \begin{table}
  \caption{Quark model charges. The superscripts denote scalar,
pseudoscalar, vector-isoscalar, and vector-isovector quark charges,
respectively.}
   \begin{tabular}{ccccc}
    & $g^{\rm s}$ & $g^{5}$ & $g^{+}$ & $g^{-}$ \\
     \tableline
      $u$   &  $+1/3$  &  $+1/5$  &  $+1/3$  &  $+1$  \\
      $d$   &  $+1/3$  &  $-1/5$  &  $+1/3$  &  $-1$  \\
   \end{tabular}
  \label{tableone}
 \end{table}

\mediumtext
 \begin{table}
  \caption{Proton, neutron, isoscalar, and isovector
           meson-nucleon coupling constants.}
   \begin{tabular}{ccccccc}
     & $g^{\sigma}$ & $g^{\pi}$
     & $g^{\omega}$ & $f^{\omega}$
     & $g^{\rho}  $ & $f^{\rho}  $                       \\
     \tableline
      $p$  &  $+1$  &  $+1$
           &  $+1$  &  $0 $
           &  $+1$  &  $+4$                              \\
      $n$  &  $+1$  &  $-1$
           &  $+1$  &  $0 $
           &  $-1$  &  $-4$                              \\
      $0$  &  $+1$  &  ${3\over 10}{\Delta m\over m}$
           &  $+1  $  &  $0$
           &  $0   $  &  ${3\over 2}{\Delta m\over m}$   \\
      $1$  &  $0$  &  $+1$
           &  $0   $  &  ${5\over 6}{\Delta m\over m}$
           &  $1   $  &  $+4$                            \\
   \end{tabular}
  \label{tabletwo}
 \end{table}

\mediumtext
 \begin{table}
  \caption{Meson masses, coupling constants, tensor-to-vector ratio,
           and cutoff parameters of the Bonn B potential.}
   \begin{tabular}{ccccc}
    Meson    &  Mass (MeV) & $g^{2}/4\pi$  & $f/g$ & $\Lambda$~(MeV) \\
     \tableline
    $\pi$    &     138     &     14.21     &  ---  &      1700       \\
    $\eta$   &     549     &      2.25     &  ---  &      1500       \\
    $\rho$   &     769     &      0.42     &  6.1  &      1850       \\
    $\omega$ &     783     &     11.13     &  0.0  &      1850       \\
   \end{tabular}
  \label{tablethree}
 \end{table}

\mediumtext
 \begin{table}
  \caption{Spin singlet-triplet mixing angles
           $\bar{\gamma}_{\lower 2pt \hbox{$\scriptstyle J$}}$
           (in degrees) at a laboratory energy of
           $T_{\rm lab}=183$~MeV. The values in parenthesis
           use a form factor computed from the quark model
           (see text for details).}
   \begin{tabular}{cccc}
     $J$     & $\langle\rho|H|\omega\rangle|_{\rm on}$       &
     $\langle\rho|H|\omega\rangle|_{\rm off}+(\omega-\rho$)  &
     $\langle\rho|H|\omega\rangle|_{\rm on }+(\omega-\rho$)  \\
     \tableline
      1 & $3.41\times10^{-2}$ & $3.32\times10^{-2}$ ($3.66\times10^{-2}$) &
                                $6.70\times10^{-2}$ ($7.22\times10^{-2}$) \\
      2 & $4.51\times10^{-2}$ & $4.40\times10^{-2}$ ($4.88\times10^{-2}$) &
                                $8.77\times10^{-2}$ ($9.52\times10^{-2}$) \\
      3 & $3.77\times10^{-3}$ & $3.71\times10^{-3}$ ($4.71\times10^{-3}$) &
                                $6.00\times10^{-3}$ ($7.55\times10^{-3}$) \\
      4 & $8.04\times10^{-4}$ & $8.58\times10^{-4}$ ($1.12\times10^{-3}$)&
                                $1.18\times10^{-3}$ ($1.55\times10^{-3}$)\\
   \end{tabular}
  \label{tablefour}
 \end{table}

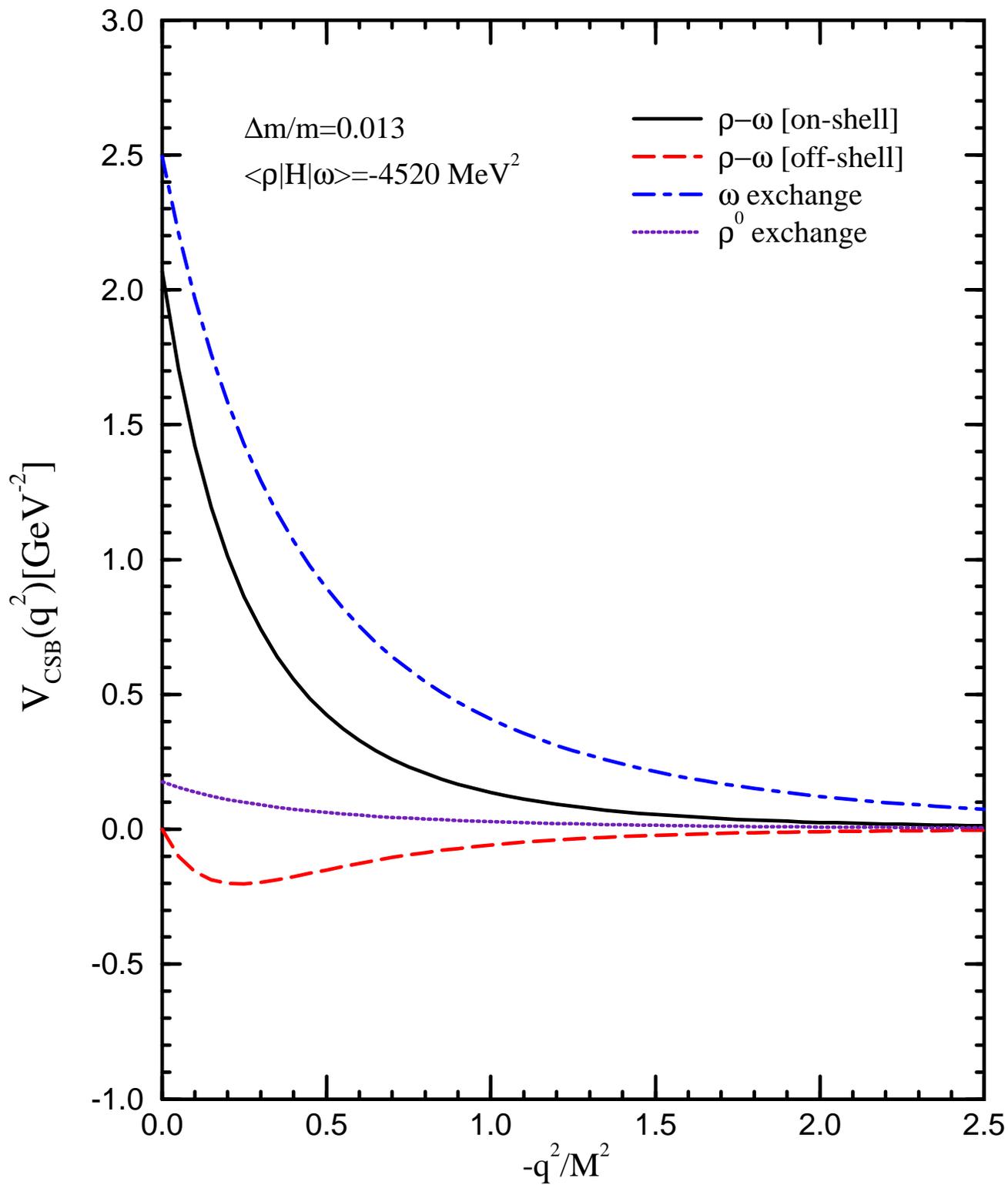
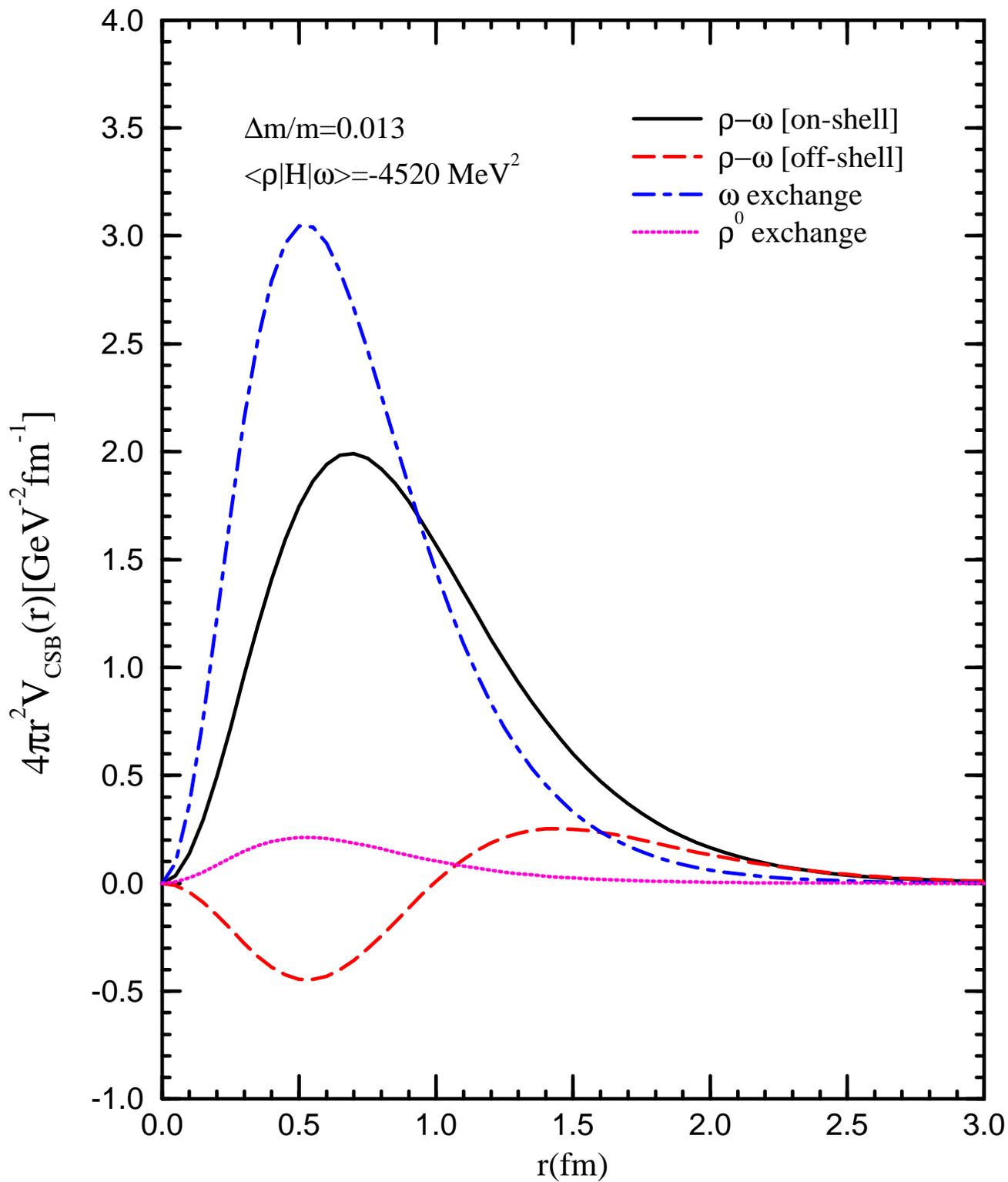
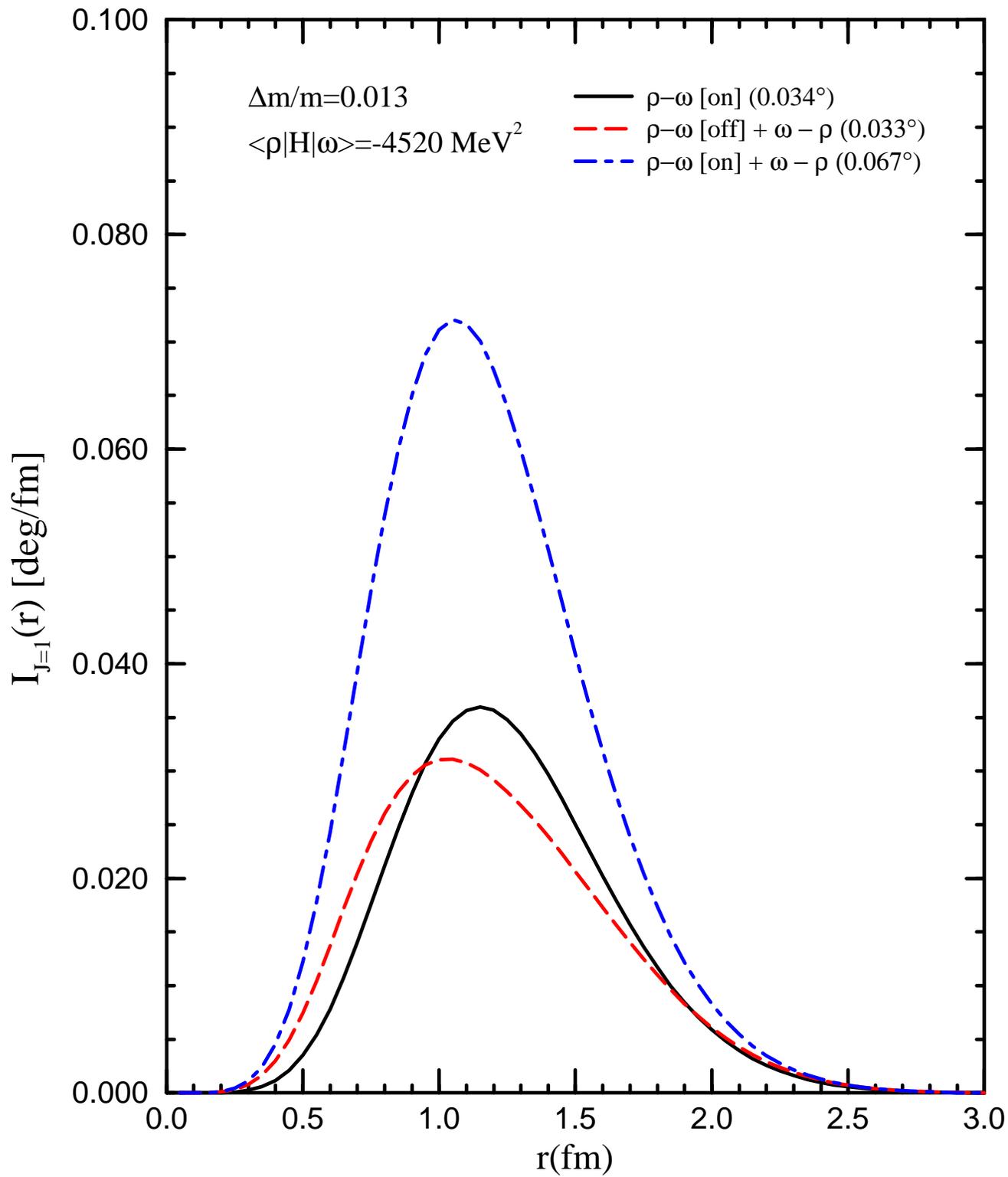
\end{document}